\RequirePackage{fix-cm}
\documentclass[smallextended,natbib]{svjour3}

\usepackage{graphicx}
\usepackage{multirow}
\usepackage{amsmath,amssymb,amsfonts}
\usepackage{mathrsfs}
\usepackage{xcolor}
\usepackage{textcomp}
\usepackage{manyfoot}
\usepackage{booktabs}
\usepackage{algorithm}
\usepackage{algorithmicx}
\usepackage{algpseudocode}
\usepackage{listings}

\usepackage{mathtools}
\usepackage{empheq}
\usepackage{enumitem}
\usepackage{threeparttable}
\usepackage{comment}

\DeclareMathOperator*{\argmin}{arg\,min}

\smartqed
\begin{document}

\title{An Algebraic Approach to Constrained K-means Clustering}
\titlerunning{An Algebraic Approach to Constrained K-means Clustering}

\author{Hiroki Oohori \and Kentaro Fukushima \and Michio Yamamoto}
\authorrunning{Oohori et al.}

\institute{
Hiroki Oohori \and Kentaro Fukushima \and Michio Yamamoto
\at Graduate School of Human Sciences, The University of Osaka, Osaka, Japan
\and
Kentaro Fukushima
\at Graduate School of Education, The University of Tokyo, Tokyo, Japan
\and
Michio Yamamoto
\at Center for Advanced Intelligence Project, RIKEN, Tokyo, Japan
\and
Michio Yamamoto
\at Data Science and AI Innovation Research Promotion Center, Shiga University, Shiga, Japan
}
\date{}

\maketitle

\begin{abstract}
    Clustering is typically performed using only object features, although additional information about relationships between objects may be available and can be incorporated through constraints. 
    Among these, pairwise constraints require two objects to belong to the same cluster or to different clusters. Several extensions of the k-means algorithm accommodate such constraints, but they sometimes fail to find a feasible assignment or incur high computational cost. Motivated by these limitations, we propose an algebraic approach to k-means clustering with pairwise constraints. The algorithm simplifies the constraints using Gr\"{o}bner bases before optimization, improving the efficiency of cluster assignment. It can also detect whether the constraint set has a feasible assignment before optimization. Comparative experiments with existing algorithms showed that the proposed algorithm achieved comparable or better clustering quality at a lower computational cost, particularly when the numbers of constraints and clusters were small.

\end{abstract}

\section{Introduction}
\label{sec_1}
Clustering algorithms generally rely only on the features of each object and do not use information indicating how the objects should be assigned to clusters.
However, knowledge or information related to the true cluster but not represented in features is often available.
It is known that such background information or expert knowledge can improve clustering performance \citep{wagstaff2001constrained}.
One approach is to use constrained clustering and treat such information as constraints.
Pairwise constraints are among the most extensively studied types of constraints in constrained clustering.
There are two types of pairwise constraints: must-link (ML) constraints and cannot-link (CL) constraints.
An ML constraint requires a pair of objects to be assigned to the same cluster.
Conversely, a CL constraint requires a pair of objects to be assigned to different clusters.
Clustering with pairwise constraints has not only been studied theoretically but also applied to practical problems in fields such as metagenomics, remote sensing, and digital soil mapping \citep{pan2022deep,qin2022active,zhu2025using}.

When considering pairwise constraints, whether constraint violations are permitted is a key issue.
For the case without constraint violations, many simple and fundamental algorithms have been developed \citep{gonzalez2025semi,piccialli2022exact,mansueto2025efficiently}.
Also, algorithms permitting constraint violations have been developed.
Such algorithms are especially effective when the background knowledge and the resulting constraints are not completely reliable \citep{basu2004active,bilenko2004integrating,gonzalez2020dils,baumann2022k}.
In this setting, it is often necessary to consider how strongly constraint violations should be penalized or how reliable each constraint is.
These are often represented through hyperparameters for penalty terms or weights for individual constraints, which must be determined before clustering.
In this study, we consider the setting without any constraint violations to make the setting simpler, reduce user decisions, and keep the proposed algorithm basic as a benchmark for algebraic approaches.

In developing such constraint-satisfying algorithms, the number of objects is often a concern, especially when this number is large \citep{chumpitaz2026pass}.
One way to address this large-scale issue is to extend the k-means algorithm.
K-means is an unconstrained clustering algorithm that iteratively updates the centroid of each cluster and the assignments of the objects \citep{macqueen1967multivariate}.
Because of its computational efficiency, k-means provides a reasonable basis for clustering large-scale data \citep{baumann2022pccc}.

COPK-means (COnstrained Partitioning variant of K-means; \citealp{wagstaff2001constrained}) is one of the best-known extensions of k-means for settings in which all constraints must be satisfied.
COPK-means retains the iterative assignment-and-update framework of k-means while incorporating pairwise constraints into the assignment step.
However, COPK-means can fail to reach an optimal assignment and may not even find a feasible assignment because its assignments depend on the order in which objects are assigned.
BLPKM (Binary Linear Programming-based K-Means; \citealp{baumann2020binary}) addresses these problems by formulating the assignment problem as a binary linear programming problem and assigning all objects simultaneously.
However, it can incur high computational costs \citep{mansueto2025efficiently}.

In this study, we propose a new algebraic approach to k-means clustering with pairwise constraints for large-scale data.
The proposed algorithm represents the constraints algebraically and uses Gr\"{o}bner bases \citep{buchberger1976theoretical,cox1997ideals} to obtain an equivalent representation of the constraint system that can be exploited in cluster assignment.
This algebraic formulation serves two main roles.
First, the formulation enables the systematic exclusion of infeasible candidate assignments during the search for an optimal feasible assignment, thereby reducing the computational burden of the assignment step.
This is particularly important for k-means and its extensions, which repeatedly alternate between cluster assignment and centroid update and are often run with multiple initializations because of their dependence on initial values  \citep{jain2010data}.
Second, the Gr\"{o}bner basis provides an algebraic means of determining whether any feasible cluster assignment exists under the given constraints before the iterative clustering procedure begins.
Thus, the proposed approach uses a unified algebraic framework both to facilitate constrained cluster assignment and to identify infeasible constraint sets.

The remainder of this paper is organized as follows.
Section~\ref{sec_2} introduces the notations used in this paper, reviews related studies, and briefly discusses the limitations of the algorithms.
Section~\ref{sec_3} describes the proposed algorithm, and Section~\ref{sec_4} illustrates its behavior using real data.
Section~\ref{sec_5} presents the design and results of the comparative experiments.
Finally, Section~\ref{sec_6} discusses the findings, limitations, and directions for future research.

\section{Related work}
\label{sec_2}
\subsection{Problem settings}
\label{sec_2.1}
The proposed algorithm is based on the k-means algorithm and performs cluster assignment without violating the given constraints.
Existing algorithms within this framework include COPK-means \citep{wagstaff2001constrained} and BLPKM \citep{baumann2020binary}.
We first introduce the notation and basic settings common to COPK-means, BLPKM, and the proposed algorithm.

We consider the partitioning of $n$ objects into $k$ clusters.
Each object \(i\) has its feature vector $y_i\in\mathbb{R}^d$, and each cluster $j$ is associated with a centroid $c_j\in\mathbb{R}^d$. 
Here, \(d_{ij}\) denotes the cost of assigning object \(i\) to cluster \(j\).
In this study, \(d_{ij}\) is given by the squared Euclidean distance between \(y_i\) and \(c_j\).
Let $z_{ij}\in\{0,1\}$ be an indicator variable representing whether object $i$ belongs to cluster $j$:
$z_{ij}=1$ if $i$ belongs to cluster $j$, and $z_{ij}=0$ otherwise.
We assume that ML and CL constraints are imposed on several pairs of objects.
Hereafter, we represent a constraint between objects $i$ and $i'$ as $(i,i')$, and the sets of ML and CL constraints are denoted by
$
\mathcal{E}_{ML},\mathcal{E}_{CL}
\subseteq
\{1,\ldots,n\}\times\{1,\ldots,n\},
$
respectively.

Under this setting, constrained k-means clustering proceeds as follows:
(i) initialize the cluster centroids;
(ii) assign the objects to clusters without violating the constraints in
$\mathcal{E}_{ML}$ and $\mathcal{E}_{CL}$;
(iii) update each cluster centroid to the mean of the objects assigned to that cluster; and
(iv) repeat steps (ii) and (iii) until convergence.
Because the solutions depend on the initial centroids, the algorithm is typically run with multiple initializations, and the resulting solutions are compared.
In this paper, we regard all steps except for the assignment step (ii) as common to all algorithms.
Therefore, the following subsection and Section~\ref{sec_3} focus on step (ii), in which the algorithms differ.
\subsection{Existing constrained k-means models}
\label{sec_2.2}
\paragraph{COPK-means}
A representative extension of the k-means algorithm incorporating pairwise constraints is COPK-means \citep{wagstaff2001constrained}.
Its basic framework is the same as that of the k-means algorithm: the assignment and update steps are repeated.
In the assignment step, however, each object is assigned to the nearest cluster among those that satisfy all constraints.

\begin{figure}[t]
  \centering
  \includegraphics[width=0.95\linewidth]
  {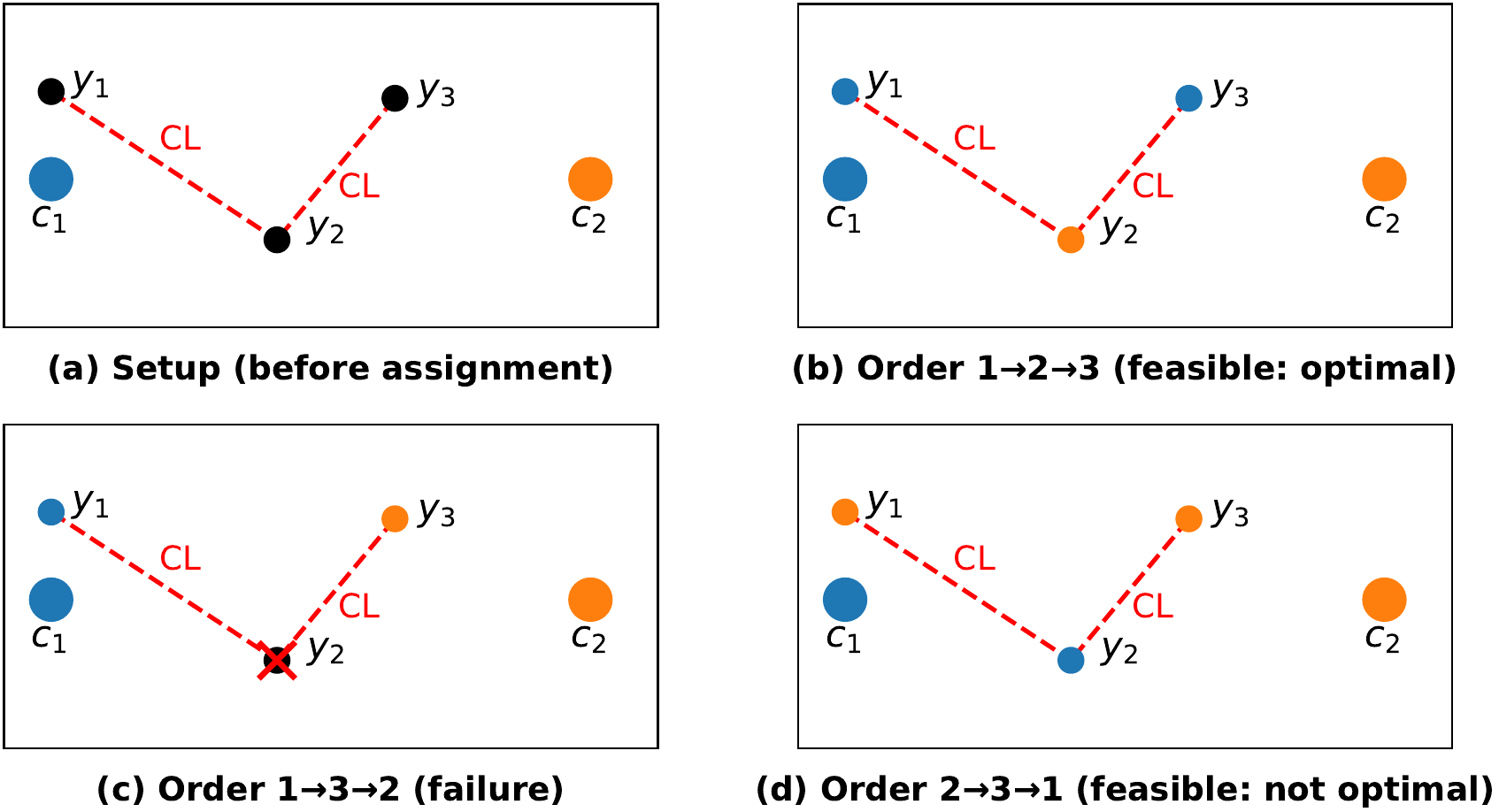}
  \caption{Order dependence in COPK-means. Panel (a) is an illustration of the setting.
    There are two clusters with centroids $c_1$ and $c_2$.
    The points of objects in feature space are $y_1$, $y_2$, and $y_3$, and a CL constraint is imposed between \(1\) and \(2\) and another between \(2\) and \(3\).
    Panels (b), (c), and (d) illustrate the order dependence of COPK-means.
    Colors of points represent clusters to which objects are assigned, and the numbers in parentheses indicate the order in which the objects are assigned.
    In (b), COPK-means reaches the optimal assignment; in (c), it cannot assign object $2$; and in (d), it obtains another, nonoptimal assignment.}
  \label{fig:cop-order}
\end{figure}

In this algorithm, assignments depend on the order in which the objects are processed, and a feasible assignment may not be found even when one exists.
These problems are illustrated in Figure~\ref{fig:cop-order}.
As shown in Figure~\ref{fig:cop-order}(a), consider assigning objects
$1,2,3$, represented by their feature vectors $\{y_1,y_2,y_3\}$,
to two clusters,
$\mathcal{C}_1$ and $\mathcal{C}_2$, subject to $(1,2),(2,3)\in\mathcal{E}_{CL}$.
If the objects are assigned in the order
$1\rightarrow 2\rightarrow 3$,
a feasible and optimal assignment (i.e., one with the smallest total assignment cost) can be obtained, as shown in Figure~\ref{fig:cop-order}(b).
However, as shown in Figure~\ref{fig:cop-order}(c), if they are assigned in the order
$1\rightarrow 3\rightarrow 2$,
$1$ and $3$ are assigned to
$\mathcal{C}_1$ and $\mathcal{C}_2$, respectively.
Consequently, $2$ cannot be assigned to either cluster because it cannot belong to the same cluster as either \(1\) or \(3\), and the algorithm returns ``FAIL.'' 
Furthermore, as shown in Figure~\ref{fig:cop-order}(d), if the assignment order is
$2\rightarrow 3\rightarrow 1$,
$2$ is assigned to $\mathcal{C}_1$, after which $1$ and $3$ are assigned to $\mathcal{C}_2$.
The resulting assignment is not optimal.

\paragraph{BLPKM}
Another extension of k-means for pairwise constraints is BLPKM \citep{baumann2020binary}.
This algorithm assigns objects by solving the following binary linear programming problem with respect to
$Z=\{z_{ij}\mid i=1,\ldots,n,\ j=1,\ldots,k\}$.

\begin{empheq}{align}
\min_{Z}\;&
\sum_{i=1}^{n}\sum_{j=1}^{k} d_{ij}z_{ij},
\label{Lossfunc}\\
\text{s.t.}\;&
\sum_{j=1}^{k} z_{ij}=1,
&& (i=1,\ldots,n)
\label{eq:blpkm_c1}\\
&
\sum_{i=1}^{n} z_{ij}\geq 1,
&& (j=1,\ldots,k)
\label{eq:blpkm_c2}\\
&
z_{ij}=z_{i'j},
&&
\bigl((i,i')\in\mathcal{E}_{ML};\ j=1,\ldots,k\bigr)
\label{eq:blpkm_c3}\\
&
z_{ij}+z_{i'j}\leq 1,
&&
\bigl((i,i')\in\mathcal{E}_{CL};\ j=1,\ldots,k\bigr).
\label{eq:blpkm_c4}
\end{empheq}
The objective function in (1) minimizes the total assignment cost. Constraints (2) and (3) ensure that each object is assigned to exactly one cluster and that every cluster contains at least one object, respectively. Constraints (4) and (5) encode the ML and CL constraint requirements, respectively.

Because BLPKM determines the assignments of all objects simultaneously, it avoids the order dependency.
However, it requires a binary linear programming problem to be solved every time the assignments are updated.
Consequently, its iterative optimization can impose a substantial computational burden \citep{mansueto2025efficiently}.

\section{Proposed algorithm}
\label{sec_3}
To address the limitations of the existing algorithms, we propose the assignment algorithm shown in Algorithm~\ref{alg:proposed}.
The proposed algorithm consists of a preprocessing step and an assignment step.
The preprocessing step needs to be performed only once before the iterations, while the assignment step is repeated as in the k-means algorithm.
In Algorithm~\ref{alg:proposed},
\(\mathcal{G}_{ML}\) and \(\mathcal{G}_{CL}\) are undirected graphs whose vertex set is \(\{1,\ldots,n\}\) and whose edge sets are \(\mathcal{E}_{ML}\) and \(\mathcal{E}_{CL}\), respectively.
The algorithm also uses reduced Gr\"{o}bner bases, whose definition and basic properties are provided in \ref{Append_GB}.

\begin{algorithm}[t]
\caption{Proposed assignment algorithm}
\label{alg:proposed}
\begin{algorithmic}[1]

\Require
\(\{y_1,\ldots,y_n\},\ k,\ d_{ij},\ \mathcal{E}_{ML},\ \mathcal{E}_{CL},\ \mathcal{G}_{ML},\ \mathcal{G}_{CL}\)

\Statex \textbf{Preprocessing step}

\State
\(\{B_1,\ldots,B_P\}\gets\) the vertex sets of the connected components of
\(\mathcal{G}_{ML}\)

\State
\(x_{pj}\in\{0,1\}\) represents whether block \(B_p\) is assigned to cluster \(j\)

\State
\(\tilde{\mathcal{E}}_{CL}\gets
\{(p,p')\mid \exists(i,i')\in\mathcal{E}_{CL};\ i\in B_p,\ i'\in B_{p'}\}\)

\If{\((p,p)\in\tilde{\mathcal{E}}_{CL}\) for some \(p\)}
\State \Return \textbf{FAIL}
\EndIf

\State
Define \(\tilde{\mathcal{G}}_{CL}\) as the undirected graph with vertex set
\(\{1,\ldots,P\}\) and edge set \(\tilde{\mathcal{E}}_{CL}\)

\State
\(\forall p;\ x_{pk} \coloneqq 1-\sum_{j=1}^{k-1}x_{pj}\)

\State
\(\mathcal{O}\gets
\{p\mid \exists p';\ (p,p')\in\tilde{\mathcal{E}}_{CL}\}\)

\State
\(\{O_1,\ldots,O_S\}\gets\) the vertex sets of the connected components of
\(\tilde{\mathcal{G}}_{CL}\) induced by \(\mathcal{O}\)

\For{\(s=1,\ldots,S\)}
    \State
    \(X_s\gets\{x_{pj}\mid p\in O_s,\ j=1,\ldots,k\}\)

    \State
    \(P_s\gets
    \{x_{pj}x_{p'j}\mid
    (p,p')\in\tilde{\mathcal{E}}_{CL},\
    p,p'\in O_s,\ j=1,\ldots,k\}\)

    \State
    \(G_s\gets\) the reduced Gr\"{o}bner basis of \(\langle P_s\rangle\)

    \If{\(G_s=\{1\}\)}
    \State \Return \textbf{FAIL}
    \EndIf

    \State
    \(V_s\coloneqq
    \{X_s\mid
    \forall g_t^{(s)}\in G_s;\
    g_t^{(s)}(X_s)=0\}\)
\EndFor

\Statex \textbf{Assignment step}

\State
\(\forall p,\forall j;\
D_{pj}\gets\sum_{i\in B_p}d_{ij}\)

\State
\(\forall p\notin\mathcal{O};\
\hat{x}_{pj}\gets
\begin{cases}
1 & \left(j=\argmin_{j'=1,\ldots,k}D_{pj'}\right),\\
0 & (\text{otherwise})
\end{cases}\)

\State
\(\forall O_s;\
\hat{x}_s\gets
\argmin_{X_s\in V_s}
\sum_{p\in O_s}\sum_{j=1}^{k}x_{pj}D_{pj}\)

\State
\(\forall p\in O_s;\
\hat{x}_{pj}\gets(\hat{x}_s)_{pj}\)

\State
\(\forall i\in B_p;\
z_{ij}\gets\hat{x}_{pj}\)

\State
\Return
\(\{z_{ij}\mid i=1,\ldots,n,\ j=1,\ldots,k\}\)
or \textbf{FAIL}

\end{algorithmic}
\end{algorithm}
In the preprocessing step, we first use the Union--Find algorithm \citep{huang2023simpler} to group the objects connected by paths in
\(\mathcal{G}_{ML}\) into blocks \(B_p\) (\(p=1,\ldots,P\)).
Because all objects in a block \(B_p\) must belong to the same cluster, the subsequent procedure only needs to determine the cluster membership of each block.
Correspondingly, let \(x_{pj}\) (\(p=1,\ldots,P\), \(j=1,\ldots,k\)) be an indicator variable representing whether block \(B_p\) belongs to cluster \(j\), in the same manner as \(z_{ij}\).
Note that the indicator for the \(k\)th cluster is determined by the other indicators because
$
x_{pk}
=
1-\sum_{j=1}^{k-1}x_{pj}.
$
Afterwards, the CL constraint, originally defined between objects, is converted into a constraint between the blocks containing those objects.
The resulting set of block-level CL constraints is
$
\tilde{\mathcal{E}}_{CL}
=
\left\{
(p,p')
\;\middle|\;
\exists(i,i')\in\mathcal{E}_{CL}
\text{ such that }
i\in B_p,\ i'\in B_{p'}
\right\}.
$
We also define the undirected graph \(\tilde{\mathcal{G}}_{CL}\), obtained from \(\mathcal{G}_{CL}\) by representing each block index in \(\{1, \ldots, P\}\) as a vertex, with edges given by \(\tilde{\mathcal{E}}_{CL}\).
If a CL constraint exists between two objects in the same block \(B_p\), then
\((p,p)\in\tilde{\mathcal{E}}_{CL}\).
In this case, the constraints are infeasible, and the algorithm returns ``FAIL.''

Additionally, we define 
$
\mathcal{O}
=
\left\{
p
\;\middle|\;
\exists p'
\text{ such that }
(p,p')\in\tilde{\mathcal{E}}_{CL}
\right\},
$
which consists of the blocks involved in at least one CL constraint.
The set \(\mathcal{O}\) is partitioned into \(O_1,\ldots,O_S\) according to the connected components of \(\tilde{\mathcal{G}}_{CL}\); namely, \(p\) and \(p'\) belong to the same \(O_s\) if and only if there exists a path between them in \(\tilde{\mathcal{G}}_{CL}\). The assignments of blocks belonging to the same \(O_s\) affect one another through the CL constraints. By contrast, the assignment of a block belonging to an \(O_s\) is not affected by blocks belonging to another \(O_{s'}\) or by blocks not included in \(\mathcal{O}\).
Therefore, we focus on each \(O_s\) and consider the assignments of its blocks.
For convenience, let
$
X_s
=
\left\{
x_{pj}
\;\middle|\;
p\in O_s,\ j=1,\ldots,k
\right\}
$
denote the set of indicator variables associated with \(O_s\).
For each \(O_s\), the proposed algorithm next computes a reduced Gr\"{o}bner basis from the constraint polynomials.
Specifically, for each \((p,p')\in\tilde{\mathcal{E}}_{CL}\), the CL constraint for cluster \(j\) is represented by
$
x_{pj}x_{p'j}=0.
$
To consider these constraints, let
$
P_s
=
\left\{
x_{pj}x_{p'j}
\;\middle|\;
(p,p')\in\tilde{\mathcal{E}}_{CL},\
p,p'\in O_s,\
j=1,\ldots,k
\right\}
$
be the set of these constraint polynomials for \(O_s\), where
$
x_{pk}=1-\sum_{j=1}^{k-1}x_{pj}.
$
Let \(G_s\) be the reduced Gr\"{o}bner basis of the ideal \(\langle P_s\rangle\).
Since \(G_s\) generates the same ideal as \(P_s\) \citep{cox1997ideals}, replacing \(P_s\) with \(G_s\) does not change the set of common solutions.
We then define the set of feasible solutions as
\[
V_s
\coloneqq
\left\{
X_s
\;\middle|\;
\forall g_t^{(s)}\in G_s;\ g_t^{(s)}(X_s)=0
\right\}.\]
By Theorem~\ref{nullstellensatz}, 
if \(G_s=\{1\}\), the corresponding polynomial system has no solution, and hence no assignment satisfies all the constraints.
The algorithm therefore returns ``FAIL'' before the iterative assignment procedure begins.

In the assignment step, the optimal assignment for each \(O_s\) is searched for using a branch-and-bound algorithm \citep{land2009automatic}.
During the search, each candidate assignment is examined to determine whether it can be optimal and whether it violates the constraints.
Notably, the latter can be detected using \(G_s\).
The details of this search are described in~\ref{Appendix branch and bound}, and Section~\ref{sec_4} illustrates the procedure with a concrete example.

\section{Illustration}
\label{sec_4}
This section illustrates the behavior of the proposed algorithm using a real dataset.
Subsection~\ref{sec_feadible} considers a feasible constraint set, and Subsection~\ref{sec_infeasible} considers an infeasible constraint set.

In this section, we use the Haberman's Survival dataset \citep{haberman's_survival_43}.
The dataset contains data on patients who underwent surgery for breast cancer and classifies them according to postoperative survival.
Class 1 consists of patients who survived for at least 5 years after surgery, whereas class 2 consists of patients who died within 5 years.
The features are the patient's age at the time of surgery, the year of surgery represented by the last two digits of the year, and the number of positive axillary lymph nodes detected.
We extract 10 objects from the dataset.
The extracted data are shown in Table~\ref{tab:surval data}.

\begin{table}[t]
\caption{Extracted observations from the Haberman's Survival dataset}
\label{tab:surval data}
\centering
\begin{tabular}{ccccc}
\toprule
ID & age & operation year & positive nodes & survival status \\
\midrule
1  & 57 & 61 & 30 & 2 \\
2  & 59 & 62 & 35 & 2 \\
3  & 60 & 65 & 23 & 2 \\
4  & 30 & 65 & 0  & 1 \\
5  & 33 & 60 & 0  & 1 \\
6  & 34 & 60 & 0  & 1 \\
7  & 41 & 59 & 8  & 1 \\
8  & 41 & 69 & 8  & 1 \\
9  & 44 & 64 & 6  & 2 \\
10 & 44 & 58 & 9  & 2 \\
\bottomrule
\end{tabular}

\vspace{2mm}
\begin{minipage}{\linewidth}
\footnotesize
\raggedright
Note:
Ten observations were extracted from the Haberman's Survival dataset.
``ID'' denotes the observation identifier;
``age'' denotes the patient's age at the time of surgery;
``operation year'' denotes the year of surgery expressed as the number of years since 1900;
``positive nodes'' denotes the number of positive axillary lymph nodes detected; and
``survival status'' denotes the survival outcome
(1 = survived for at least 5 years; 2 = died within 5 years).
\end{minipage}
\end{table}

\subsection{Feasible case}
\label{sec_feadible}

\begin{figure}[t]
\centering
\includegraphics[
  width=\linewidth,
  trim=0 0 0 0,
  clip
]{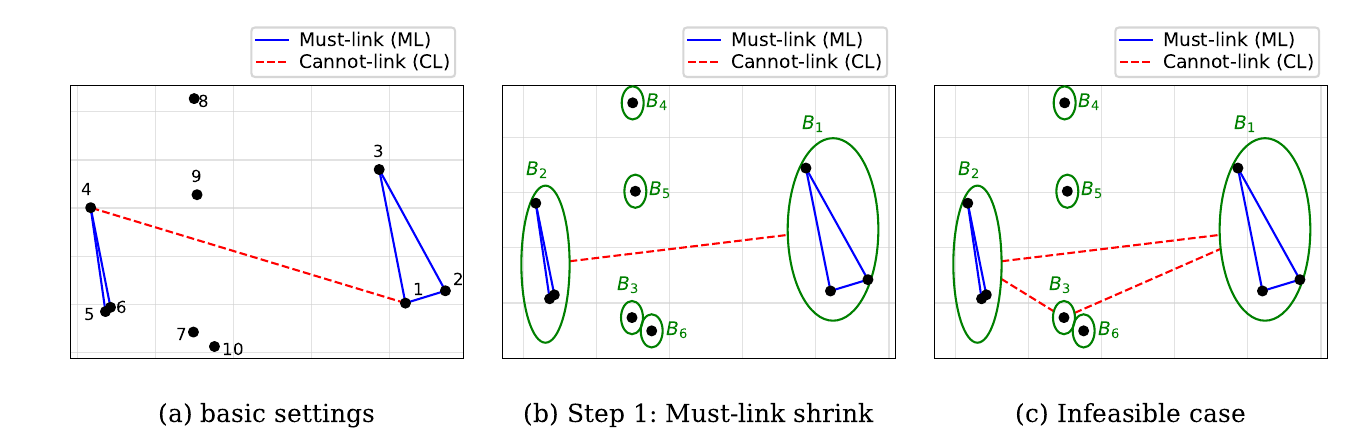}
\caption{
Principal component analysis visualization of the extracted Haberman's Survival data, with pairwise constraints overlaid.
Blue solid lines represent must-link constraints, and red dashed lines represent cannot-link constraints.
(a) Initial configuration.
(b) Blocks obtained by merging objects connected through must-link constraints.
The blocks, indicated by green ellipses, are
$B_1=\{1,2,3\}$,
$B_2=\{4,5,6\}$,
$B_3=\{7\}$,
$B_4=\{8\}$,
$B_5=\{9\}$, and
$B_6=\{10\}$.
(c) Infeasible case.
}
\label{fig:surval illustration}
\end{figure}

We first illustrate the proposed algorithm when the constraint set is feasible.
Here, we impose ML constraints on every pair of objects within IDs 1, 2, and 3 and within IDs 4, 5, and 6 and a CL constraint between IDs 1 and 4 for illustration.
Thus, the constraint sets are
$\mathcal{E}_{ML}
=
\{(1,2),(2,3),(1,3),(4,5),(5,6),(4,6)\}$
and
$\mathcal{E}_{CL}
=
\{(1,4)\}$.
The setting is shown in Figure~\ref{fig:surval illustration}(a) using principal component analysis (PCA) visualization.

The proposed algorithm first merges objects connected by ML constraints into blocks.
In this example, $\{1,2,3\}$ and $\{4,5,6\}$ form the blocks $B_1$ and $B_2$, respectively, as shown in Figure~\ref{fig:surval illustration}(b).
The remaining objects each form a singleton block.
The blocks involved in CL constraints are then partitioned into connected components $O_s$.
Because the only CL constraint in the present example connects $B_1$ and $B_2$, we have $\mathcal{O}=O_1$.
For comparison, if an additional CL constraint were imposed between $B_3$ and $B_4$, then
$\mathcal{O}=\{1,2,3,4\}$,
$O_1=\{1,2\}$, and
$O_2=\{3,4\}$.

The proposed algorithm computes a reduced Gr\"{o}bner basis separately for each $O_s$.
Because there are two clusters in this example, $x_{p2}=1-x_{p1}$.
The CL constraint between $B_1$ and $B_2$ is therefore represented by
$
x_{11}x_{21}=0\ \text{and}\ 
(1-x_{11})(1-x_{21})=0.
$
Accordingly,
$P_1
=
\{x_{11}x_{21},(1-x_{11})(1-x_{21})\}$.
Under the graded lexicographic order, the reduced Gr\"{o}bner basis of the ideal generated by $P_1$ is
$G_1
=
\{x_{11}+x_{21}-1,\ x_{21}^{2}-x_{21}\}$.
Afterwards, the optimal assignment is found according to the Gr\"{o}bner basis and tentative centroids, using a branch-and-bound algorithm.

\subsection{Infeasible case}
\label{sec_infeasible}
The proposed algorithm can readily detect infeasibility, that is, when no assignment satisfies all the constraints, and we next illustrate how it detects such infeasibility.
In this section, we construct infeasible cases by adding constraints to the basic setting introduced in Subsection~\ref{sec_feadible}.

Some infeasibilities can be detected immediately after objects are merged into blocks according to the ML constraints.
Such infeasibility occurs when a pair of objects with a CL constraint is also constrained by some ML constraints, resulting in the objects being required to belong to both the same and different clusters.
For example, adding a CL constraint between IDs 1 and 2 or an ML constraint between IDs 3 and 6 results in a CL constraint between two objects in the same block, which makes the constraints infeasible.

The proposed algorithm can also detect more complex infeasibilities, once the Gr\"{o}bner bases are obtained.
For example, if CL constraints are added between IDs 6 and 7 and between IDs 7 and 1 in addition to the setting in Subsection~\ref{sec_feadible}, then $\mathcal{E}_{CL}=\{(1,4),(1,7),(6,7)\}$, as shown in Figure~\ref{fig:surval illustration}(c).
We can see that this case is infeasible because \(B_1\), \(B_2\), and \(B_3\) must be assigned to different clusters with only two available clusters.
The proposed algorithm can also detect such infeasibility before the iterative optimization begins.
In this case, \(O_1=\{1,2,3\}\), and the corresponding set of constraint polynomials \(P_1\) is \(\{ x_{11}x_{21}, (1-x_{11})(1-x_{21}), x_{21}x_{31}, (1-x_{21})(1-x_{31}), x_{11}x_{31}, (1-x_{11})(1-x_{31}) \}.\)
The reduced Gr\"{o}bner basis of the ideal generated by \(P_1\) is \(G_1=\{1\}\). By Theorem~\ref{nullstellensatz}, this polynomial system has no common solution. Therefore, we can conclude that no assignment satisfies the constraints, and the proposed algorithm returns ``FAIL'' at this point.

\section{Experiments}
\label{sec_5}
We compared the proposed algorithm with BLPKM, COPK-means, and ordinary (unconstrained) k-means through numerical experiments.
The experiments were designed to evaluate the effectiveness and limitations of the proposed algorithm.
Because the proposed algorithm is intended for large-scale applications, we used datasets containing approximately $n=10{,}000$ objects.

All experiments were conducted on a computer running Windows 11 and equipped with an Intel Core i9-10900X processor at 3.70 GHz and 255.68 GB of RAM.
Python 3.12.7 was used for implementation.
We also used Groebner.jl 0.10.3 \citep{demin2024groebnerjl} in Julia 1.12.6 \citep{bezanson2017julia} to compute Gr\"{o}bner bases and Gurobi Optimizer 13.0.1 \citep{gurobi} to solve binary linear programming problems.

\subsection{Datasets}
\begin{figure}
    \centering
    \includegraphics[width=0.95\linewidth]{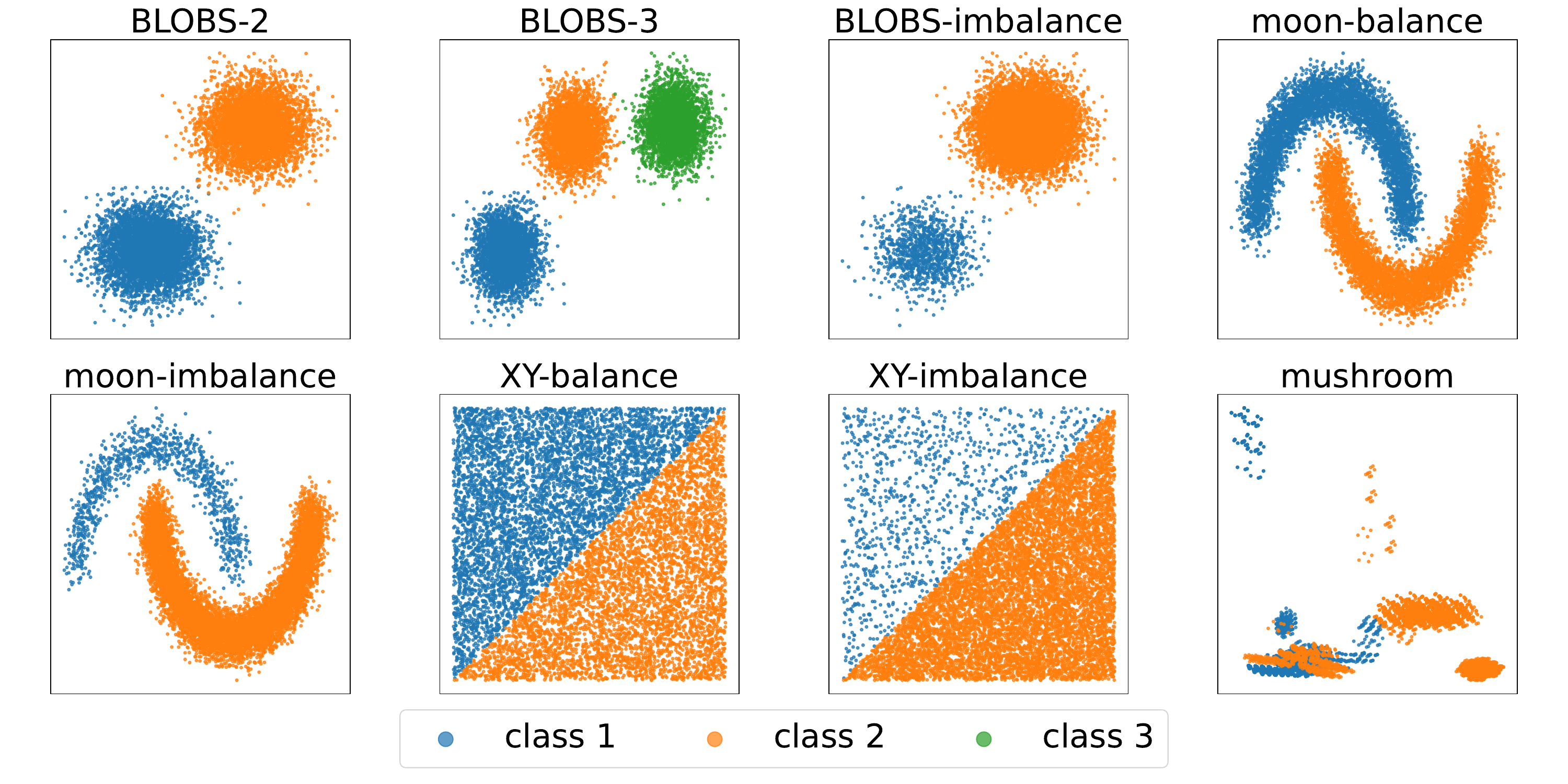}
    \caption{Scatter plots of the datasets, color-coded by class}
    \label{fig: datasets}
\end{figure}

We used the eight datasets listed in Table~\ref{table:datasets}.
The datasets were fixed throughout the experiments, whereas multiple constraint sets were independently generated for each dataset, as described in Section~\ref{subsec_experiment}.
Figure~\ref{fig: datasets} shows scatter plots of the datasets.
For every dataset, each feature was standardized before clustering; therefore, the figure presents the standardized data.
The mushroom data are projected onto two dimensions by PCA in the figure.

\begin{table}[t]
\caption{Datasets used in experiments}\label{table:datasets}%
\centering
\normalsize
\setlength{\tabcolsep}{12pt}
\renewcommand{\arraystretch}{0.9}
\begin{tabular}{@{}l c c l@{}}
\toprule
Dataset ID & $k$ & Features & Cluster size \\
\midrule
BLOBS-2 & 2 & 2 & 5000/5000 \\
BLOBS-3 & 3 & 2 & 3334/3333/3333 \\
BLOBS-imbalance & 2 & 2 & 1000/9000 \\
moon-balance & 2 & 2 & 5000/5000 \\
moon-imbalance & 2 & 2 & 1000/9000 \\
XY-balance & 2 & 2 & 5000/5000 \\
XY-imbalance & 2 & 2 & 1000/9000 \\
mushroom & 2 & 21 & 4208/3916 \\
\bottomrule
\end{tabular}
\end{table}

\paragraph{BLOBS}
The BLOBS datasets were generated using \texttt{make\_blobs} in scikit-learn \citep{scikit-learn}.
The standard deviation of each cluster was set to 1.
BLOBS-2 was used as the standard setting, and BLOBS-3 was used to examine the effect of the number of clusters.
We also used BLOBS-imbalance to examine the effect of cluster-size imbalance.

\paragraph{moon}
The moon datasets are two-dimensional, two-class datasets with nonconvex shapes generated using \texttt{sklearn.datasets.make\_moons} in scikit-learn \citep{scikit-learn}.
The Gaussian noise was set to 0.1.
To examine the effect of cluster-size imbalance, we used moon-balance, with cluster sizes of 5000/5000, and moon-imbalance, with cluster sizes of 1000/9000.

\paragraph{XY datasets}
The XY datasets were generated by sampling points $(x,y)$ from the square $[0,1]^2$.
Points satisfying $x<y$ were assigned to class 1, whereas those satisfying $x\geq y$ were assigned to class 2.
As with the moon datasets, we used a balanced dataset (XY-balance) and an imbalanced dataset (XY-imbalance).
These datasets have no clear cluster structure and are therefore difficult to handle using ordinary clustering algorithms.
Accordingly, we expected pairwise constraints to produce a greater improvement in assignment accuracy than in other datasets.

\paragraph{mushroom }
The mushroom dataset \citep{mushroom_73} is a real dataset consisting of $8{,}124$ objects.
Each object has $22$ features and belongs to one of two classes: edible or poisonous (including mushrooms of unknown edibility and those not recommended for consumption).
In Figure~\ref{fig: datasets}, the edible and poisonous classes are denoted by classes 1 and 2, respectively.
Because \textit{stalk-root} was the only feature containing missing values for multiple objects, the remaining $21$ features were used.

\subsection{Experimental design}
\label{subsec_experiment}
For each dataset, we generated the constraint sets as follows.
To construct each constraint set, $m$ distinct pairs were sampled from all possible pairs of objects.
Each sampled pair was assigned an ML constraint if the two objects had the same true label and a CL constraint otherwise.
We considered $m=100,200,500,1000$, and $2000$ and independently generated 50 constraint sets for each value of $m$.
The maximum value of $m$ was set to 2000 based on the experiments of \citet{lu2007semi}, in which the largest ratio of the number of constraints to the number of objects was approximately 0.2.
The expected proportion of ML constraints among all constraints was approximately 50\% for BLOBS-2, moon-balance, XY-balance, and mushroom; approximately 33\% for BLOBS-3; and approximately 82\% for BLOBS-imbalance, moon-imbalance and XY-imbalance.

In these experiments, each algorithm was run with 100 initializations, and an additional experiment examining the effect of the number of initializations is also reported in \ref{Appendix initialization sensitivity}.
All algorithms used the same objective function, $\sum_{j=1}^{k}\sum_{i=1}^{n} z_{ij}\|y_i-c_j\|_2^2$.
For evaluation, we measured the following criteria:
(i) Runtime, defined as the total CPU time for the 100 initializations;
(ii) Success sets, defined as the number of the 50 constraint sets for which at least one of the 100 initializations produced an assignment; and
(iii) Correct Classification Rate (CCR), defined as the proportion of agreement between the true labels and the selected assignment.
Among the 100 initializations, the assignment with the smallest objective function value was selected for calculating CCR.
For this assignment, the cluster labels were permuted to maximize agreement with the true labels.

Runtime was averaged over the 50 constraint sets.
CCR was averaged over the success sets and was treated as missing when the number of success sets was zero.
Because ordinary k-means does not depend on the constraint set, its results are represented by a single value across the different numbers of constraints for each evaluation measure.
The convergence criteria were a relative improvement of less than $10^{-4}$ and a maximum of 20 iterations.

\subsection{Results}
\begin{figure}
    \centering    \includegraphics[width=1.0\linewidth]{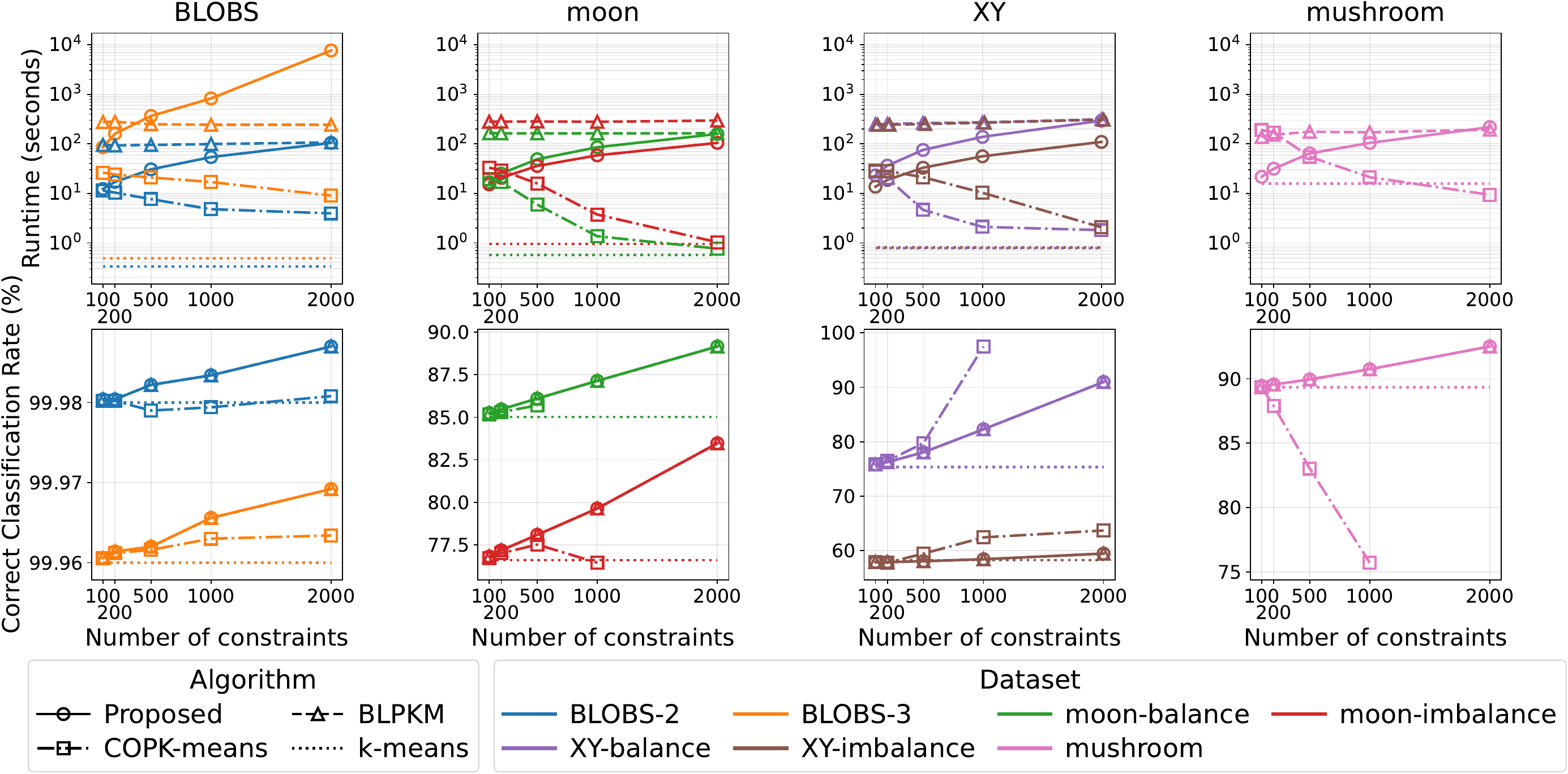}
    \caption{Runtime and Correct Classification Rate (CCR) for the four algorithms (if no assignments are obtained, the corresponding points are omitted)
}
    \label{res:runtimeCCR}
\end{figure}

Figure~\ref{res:runtimeCCR} shows the runtime and CCR results for seven datasets, and Table~\ref{res: success sets} shows the numbers of success sets.
The results for BLOBS-imbalance did not substantially affect the main findings regarding the effectiveness or characteristics of the proposed algorithm.
Therefore, for readability, the corresponding results are presented separately in~\ref{Appendix additional findings}.
Also, because the proposed algorithm, BLPKM, and ordinary k-means achieved 50 success sets under every condition, only the results for COPK-means are reported in Table~\ref{res: success sets}.

\begin{table}[t]
\caption{Success sets of COPK-means (in 50 sets)}
\label{res: success sets}%
\centering
\begin{tabular}{@{}c c c c c c c c c@{}}
\toprule
 & \multicolumn{3}{c}{BLOBS}
 & \multicolumn{2}{c}{moon}
 & \multicolumn{2}{c}{XY}
 & \\
\cmidrule(lr){2-4}
\cmidrule(lr){5-6}
\cmidrule(lr){7-8}
Number of constraints & 2 & 3 & imb & bal & imb & bal & imb & mushroom \\
\midrule
100  & 50 & 50 & 50 & 50 & 50 & 49 & 50 & 50 \\
200  & 50 & 50 & 50 & 45 & 49 & 45 & 50 & 50 \\
500  & 50 & 50 & 50 & 18 & 29 & 13 & 44 & 45 \\
1000 & 50 & 50 & 50 & 0  & 7  & 2  & 25 & 23 \\
2000 & 50 & 50 & 50 & 0  & 0  & 0  & 3  & 0  \\
\bottomrule
\end{tabular}

\begin{minipage}{\linewidth}
\footnotesize
Note: The abbreviations ``bal'' and ``imb'' denote balance and imbalance, respectively.
\end{minipage}
\end{table}

The main findings are summarized as follows.
\begin{enumerate}[label=\textit{\roman*}]
    \item When the numbers of constraints and clusters were small, the proposed algorithm was faster than BLPKM.\label{res_2}
    \item The runtime of the proposed algorithm depended more strongly on the numbers of clusters and constraints than that of BLPKM.\label{res_5}
    \item The proposed algorithm and BLPKM achieved comparable CCRs.\label{res_3}
    \item Pairwise constraints improved the CCRs of the proposed algorithm and BLPKM, although the extent of the improvement depended on the dataset and number of constraints.\label{res_4}
    \item When the number of constraints was large, COPK-means often failed to produce an assignment, whereas the proposed algorithm and BLPKM produced assignments under all conditions.\label{res_1}
\end{enumerate}

First, regarding runtime, Finding~\ref{res_2} indicates that the proposed algorithm was more efficient than BLPKM when the numbers of constraints and clusters were small.
In particular, when the number of constraints was 100, the proposed algorithm was nearly ten times faster than BLPKM for some datasets.
Therefore, the proposed algorithm is likely to be more efficient than BLPKM in applications where only a limited number of constraints are available.

By contrast, Finding~\ref{res_5} indicates one of the limitations of the proposed algorithm.
When the number of constraints increased to 2000, the runtime advantage of the proposed algorithm was no longer clear for some datasets.
Moreover, for BLOBS-3, the proposed algorithm was slower than BLPKM when the number of constraints was 500 or more.
This limitation may be related to the computation of the Gr\"{o}bner bases and the use of the branch-and-bound algorithm for optimization.
Additional experiments on these computational costs are reported in~\ref{Appendix runtime for proposed algorithm}.

Next, regarding CCR, Finding~\ref{res_3} is consistent with the fact that, except for the handling of empty clusters, the proposed algorithm and BLPKM theoretically produce the same assignments.
In particular, transforming the constraint polynomials using a Gr\"{o}bner basis does not affect the optimization problem itself.
Although slight differences in CCR were observed under some conditions, these were presumably due to numerical and implementation factors.

Furthermore, Finding~\ref{res_4} indicates that the effect of pairwise constraints on CCR differed across datasets.
This finding is consistent with related work \citep{wagstaff2001constrained}.
These dataset-specific differences are discussed further in~\ref{Appendix additional findings}.

In terms of CCR, the proposed algorithm showed an advantage over COPK-means, which often achieved lower CCRs.
In particular, for mushroom, the CCR of COPK-means decreased as the number of constraints increased.
Because mushroom is a real-world dataset, this result suggests that the performance of COPK-means may deteriorate as the number of constraints increases in some practical settings.

Finally, regarding success sets, Finding~\ref{res_1} demonstrates an advantage of the proposed algorithm over COPK-means.
The proposed algorithm and BLPKM produced assignments for all constraint sets, whereas COPK-means frequently failed to produce an assignment when the number of constraints was 500 or more.

\section{Discussion}
\label{sec_6}
In this study, we proposed an algebraic approach to constrained k-means clustering.
The proposed algorithm simplifies a system of polynomial equations equivalent to the pairwise constraints by computing a Gr\"{o}bner basis.
This simplification aims to reduce the computational cost of assignment in each iteration.
We evaluated the performance of the proposed algorithm by comparing it with existing algorithms.
The experimental results showed that the proposed algorithm can complete assignments efficiently, particularly when the numbers of constraints and clusters are small.

One limitation of this study is that the proposed algorithm handles only pairwise constraints.
Various other types of clustering constraints have been proposed \citep{gonzalez2025semi}, and extending the algebraic approach to such constraints is an important direction for future research.
For example, cardinality constraints \citep{bibi2023constrained} can specify that the number of objects assigned to cluster \(j\) must be \(n_j\), which can be represented as \(\sum_{i=1}^{n}z_{ij}=n_j\).
Besides, \(\gamma\)-constraints \citep{gonzalez2025semi} require the distance between two objects assigned to different clusters to be greater than \(\gamma\), which can be represented as \(z_{ij}z_{i'j'}=0\) for \(j\neq j'\) when \(\|y_i-y_{i'}\|_2\leq\gamma\) \citep{dao2013declarative}. 
An algebraic approach may also be effective for developing algorithms that account for these constraints.

Another limitation is that the proposed algorithm is an extension of k-means.
It is worth developing a similar algebraic approach that can be applied to other clustering algorithms or clustering-based models, such as kernel k-means and clusterwise regression.
The optimization algorithm used in the assignment step also warrants further investigation for more efficient clustering.
One possible approach is the Lasserre hierarchy \citep{lasserre2001global}.
Algorithms combining this optimization framework with Gr\"{o}bner bases have already been studied \citep{laurent2007semidefinite}, suggesting that it may provide an effective alternative optimization approach for the proposed algorithm.

\appendix
\def\thesection{Appendix~\Alph{section}}
\section{Gr\"{o}bner bases}
\label{Append_GB}

Let \(x_1,\ldots,x_n\) be variables and \(a_1,\ldots,a_n\) be nonnegative integers.
An expression \(y=\prod_{i=1}^{n}x_i^{a_i}\) is called a monomial.
Let \(U\) be a commutative ring.
A linear combination of monomials \(y_1,\ldots,y_T\) over \(U\), written as
\(p=\sum_{t=1}^{T}u_ty_t\) for some \(u_1,\ldots,u_T\in U\), is called a polynomial, and \(u_1,\ldots,u_T\) are called its coefficients.
A polynomial for which \(y_1=\cdots=y_T=1\), and hence \(p\in U\), is called a constant.

Let \(K[x_1,\ldots,x_n]\) denote the polynomial ring in \(x_1,\ldots,x_n\) over an algebraically closed field \(K\).
For \(p_1,\ldots,p_t\in K[x_1,\ldots,x_n]\), the set
\[
I
\coloneqq
\left\{
\sum_{i=1}^{t}u_ip_i
\;\middle|\;
u_i\in K[x_1,\ldots,x_n]
\right\}
\subseteq K[x_1,\ldots,x_n]
\]
is called the ideal generated by \(\{p_1,\ldots,p_t\}\) and is denoted by
\(I=\langle p_1,\ldots,p_t\rangle\).

Let \(\prec\) be a monomial order.
For a polynomial \(p\), let \(\operatorname{in}_{\prec}(p)\) denote the largest monomial appearing in \(p\) with respect to \(\prec\).
For an ideal \(I\), define
\[
\operatorname{in}_{\prec}(I)
\coloneqq
\left\langle
\left\{
\operatorname{in}_{\prec}(p)
\;\middle|\;
p\in I
\right\}
\right\rangle.
\]
A finite subset \(\{g_1,\ldots,g_s\}\subseteq I\) is called a Gr\"{o}bner basis of \(I\) with respect to \(\prec\) if
\[
\operatorname{in}_{\prec}(I)
=
\left\langle
\operatorname{in}_{\prec}(g_1),\ldots,
\operatorname{in}_{\prec}(g_s)
\right\rangle.
\]

A Gr\"{o}bner basis \(\{g_1,\ldots,g_s\}\) is called reduced if the coefficient of
\(\operatorname{in}_{\prec}(g_i)\) is \(1\) for every \(i\), and no monomial appearing in \(g_i\) is divisible by
\(\operatorname{in}_{\prec}(g_j)\) for \(i\neq j\).

For completeness, we state the weak Nullstellensatz below.
\begin{theorem}
\label{nullstellensatz}
    Let $K$ be an algebraically closed field and $I$ be an ideal of $K[x_1,\ldots,x_n]$. 
    $V(I)=\emptyset$ if and only if $I=K[x_1,\ldots,x_n]$.
\end{theorem}

\section{Branch-and-bound procedure for cluster assignment}
\label{Appendix branch and bound}
In this section, we describe in general the branch-and-bound procedure used to search for the optimal assignment for each \(O_s\), as introduced in Section~\ref{sec_3}. For this purpose, we first define the notation used in the following description. For each \(O_s\), we call a state in which the values of some of the indicator variables in \(X_s\) have been determined a partial assignment, and a state in which the values of all the indicator variables have been determined a complete assignment. For a given partial assignment, let \(\tilde{X}_s\subseteq X_s\) denote the set of indicator variables whose values have been determined. The search in the proposed algorithm starts by choosing an appropriate \(\tilde{X}_s\) and assigning values to its indicator variables. In addition, we refer to the best feasible complete assignment obtained thus far as the incumbent solution and denote its objective-function value by \(L^{(s)}_{\min}\). Using this notation, the search is performed through the following three steps.

\textbf{Step 1.}
In this step, the following two conditions are checked for a partial assignment selected arbitrarily among those not yet considered.
First, we examine whether the partial assignment can be extended to a complete assignment whose objective-function value is smaller than \(L^{(s)}_{\min}\).
Specifically, we check whether
\[
\sum_{x_{pj}\in\tilde{X}_s}x_{pj}D_{pj}
+
\sum_{
p\in
\left\{
p\in O_s
\mid
\forall j;\
x_{pj} \neq 1
\right\}
}
\min_{
\left\{
j
\mid
x_{pj}\notin\tilde{X}_s
\right\}
}
D_{pj}
\geq
L^{(s)}_{\min}.
\]
The left-hand side is a lower bound on the objective-function value of any complete assignment containing the given partial assignment.
Therefore, if this inequality holds, the partial assignment is discarded.
Second, we check whether the partial assignment violates the constraints.
Specifically, the values of the indicator variables in \(\tilde{X}_s\) are substituted into \(G_s\), and we check whether a nonzero constant appears.
If a nonzero constant appears, the partial assignment is discarded.

Partial assignments that pass these checks, that is, those that are not discarded according to either criterion, are retained as candidates that may be extended to the optimal solution.
Conversely, candidate assignments containing a discarded partial assignment are excluded from further consideration.

\textbf{Step 2.}
For each partial assignment that passes the checks in Step 1, some indicator variables are arbitrarily selected among those whose values have not yet been determined.
Then, a value of 0 or 1 is arbitrarily determined for each of them, and the checks in Step 1 are repeated.
If this extension of the partial assignment reaches a complete assignment and the complete assignment also passes the checks in Step 1, it is retained as the new incumbent solution, and \(L^{(s)}_{\min}\) is updated to its objective-function value.

\textbf{Step 3.}
Steps 1 and 2 are repeatedly applied to other partial assignments, excluding those that equal or contain a discarded partial assignment.
This iteration is repeated until all assignments for \(O_s\) either contain a discarded partial assignment or equal the incumbent solution.
Then, the incumbent solution is taken as the optimal assignment for \(O_s\).

\section{Additional findings from the main experiments}
\label{Appendix additional findings}
The experiments in Section~\ref{sec_5} focused on the findings directly related to the effectiveness and characteristics of the proposed algorithm.
Here, we present additional findings from the main experiments.

\begin{figure}[t]
    \centering
    \includegraphics[width=0.95\linewidth]{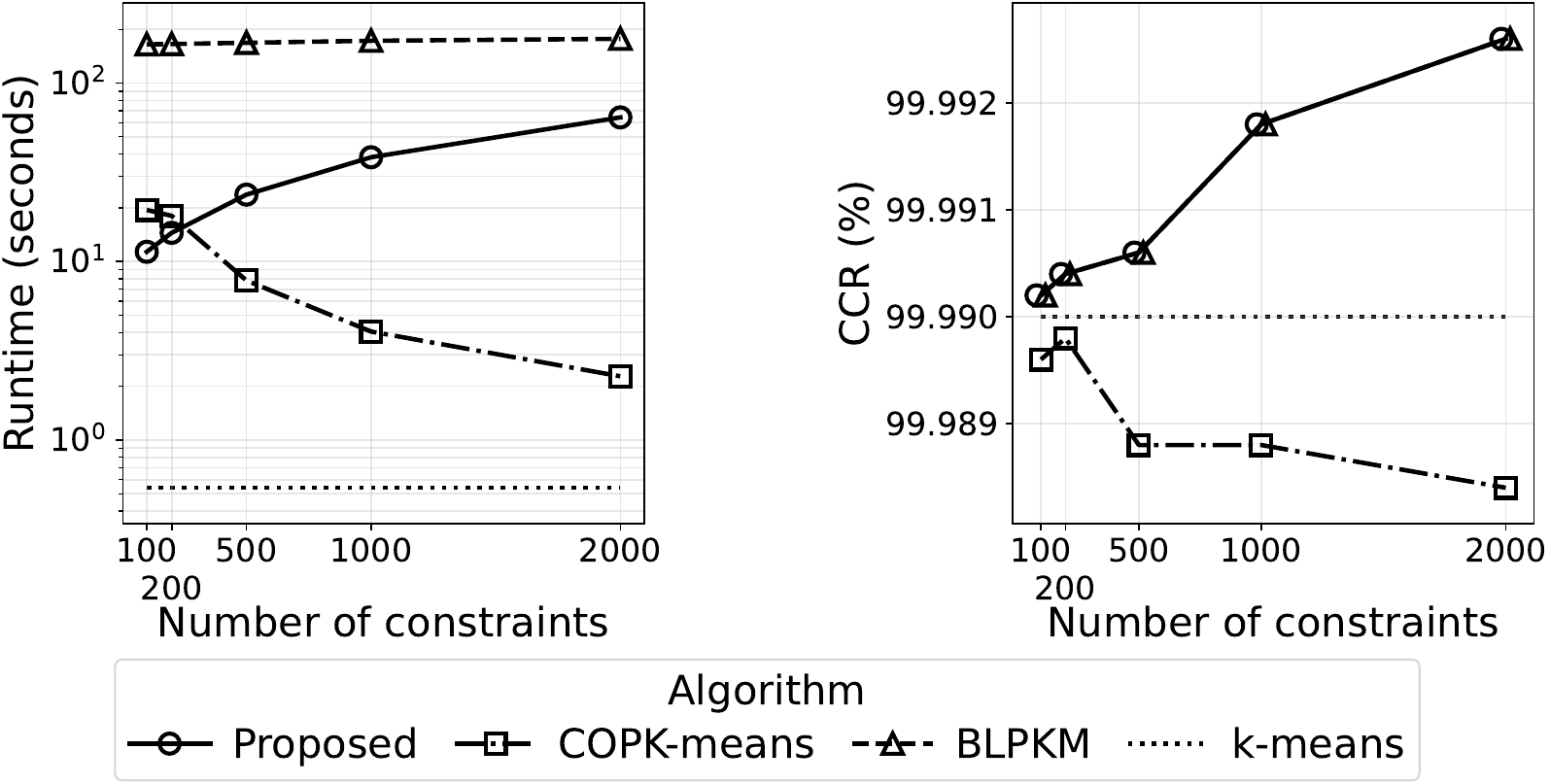}
    \caption{Runtime and Correct Classification Rate (CCR) for BLOBS-imbalance}
    \label{fig:res:BLOBS imbalance}
\end{figure}

\paragraph{BLOBS-imbalance}
The results for BLOBS-imbalance are shown in Figure~\ref{fig:res:BLOBS imbalance}.
As in the other BLOBS settings, all algorithms achieved CCRs close to 100\%.
However, the effect of cluster-size imbalance on CCR showed a different trend for BLOBS than for the moon and XY datasets.
For the moon and XY datasets, the imbalanced settings generally resulted in lower CCRs than the corresponding balanced settings.
By contrast, BLOBS-imbalance achieved slightly higher CCRs than BLOBS-2.

The slightly higher CCRs for BLOBS-imbalance may be explained by two characteristics of the BLOBS datasets.
First, almost all objects were already correctly classified in the balanced setting.
In addition, because the larger cluster in BLOBS-imbalance contained 9000 objects, its centroid was likely estimated very stably.
Consequently, these characteristics may account for the slightly higher CCRs observed for BLOBS-imbalance.

\paragraph{CCR across datasets}
The effect of pairwise constraints on CCR differed across datasets.
The BLOBS datasets were classified almost perfectly by ordinary k-means, leaving little room for improvement.
At the other extreme, ordinary k-means achieved a low CCR for XY-imbalance, and the CCRs of the proposed algorithm and BLPKM did not improve substantially even when pairwise constraints were imposed.
Larger improvements were observed for the moon datasets, for which ordinary k-means achieved moderate CCRs.
These results suggest that pairwise constraints may have limited effects when ordinary k-means already performs very well or when the dataset is not well suited to k-means-based clustering.
Accordingly, pairwise constraints may be most effective for improving cluster estimation quality when ordinary k-means achieves a moderate CCR.

The XY datasets also showed a different CCR pattern for COPK-means.
Specifically, COPK-means achieved higher CCRs than the proposed algorithm and BLPKM.
In these datasets, the within-cluster variance is large relative to the total variance, especially for XY-imbalance.
On the other hand, the objective function is designed to minimize within-cluster variation.
In this sense, the XY datasets are not well suited to these algorithms for recovering the true clusters.
Consequently, the proposed algorithm and BLPKM may have had difficulty improving CCR even when the number of constraints increased, especially for XY-imbalance.
By contrast, due to its order dependence, COPK-means may reach an assignment with a higher objective-function value.
In the XY datasets, this property may have happened to work favorably for CCR.

\section{Constraint test}
\label{Appendix; constraint test}
\begin{figure}
    \centering
    \includegraphics[width=0.4\linewidth]{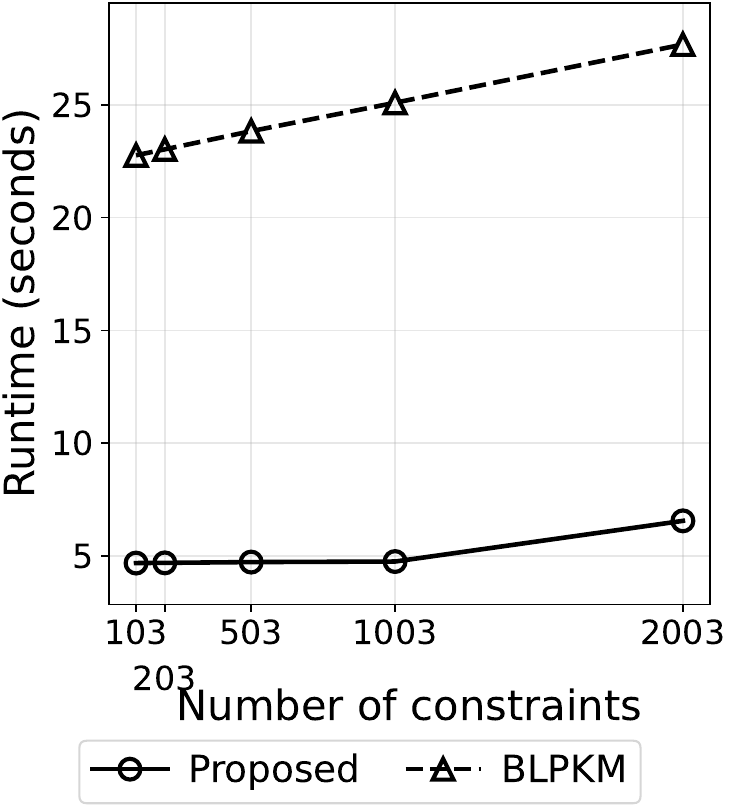}
    \caption{Runtime for infeasible settings (XY-balance)}
    \label{fig:res:infeasible}
\end{figure}

The proposed algorithm can determine whether a set of constraints is infeasible before performing cluster assignment.
To demonstrate this property, we conducted an additional experiment.
We used the XY-balance dataset, with the other experimental settings following those in Section~\ref{sec_5}.
In this experiment, we selected three objects that were not involved in any constraint and imposed CL constraints on all three pairs of these objects (thus, the total numbers of constraints were 103, 203, 503, 1003, and 2003).
Because the dataset has two clusters, no assignment of the three objects can satisfy the constraints.
We measured the runtime until execution terminated under the resulting infeasible constraint set.
Ordinary k-means performs cluster assignment without considering the constraints, and COPK-means can terminate for reasons unrelated to constraint infeasibility as described in Section~\ref{sec_2.2}.
Therefore, we report only the results for the proposed algorithm and BLPKM.
The results are shown in Figure~\ref{fig:res:infeasible}.
The results confirmed that the proposed algorithm terminated earlier than BLPKM under all conditions.
This suggests that if incorrect constraints are specified, the proposed algorithm can detect the presence of such errors faster than BLPKM.

\section{Runtime breakdown of the proposed algorithm}
\label{Appendix runtime for proposed algorithm}
As described in Section~\ref{sec_5}, the runtime of the proposed algorithm depended strongly on the number of constraints and clusters.
To investigate the source of this dependence, we measured the time required for computing the Gr\"{o}bner bases.
We regarded this time as the computational cost of Gr\"{o}bner basis computation.
Because the time required for the other operations in the preprocessing step was assumed to be negligible, we treated the difference between the total runtime and the Gr\"{o}bner basis computation time as the time required for optimization by the branch-and-bound method.
The results for BLOBS-2 and BLOBS-3 are shown in Table~\ref{Tab: runtime for proposed}.

For BLOBS-2, the Gr\"{o}bner basis computation time remained between 4.08 and 4.27 seconds for 100--1000 constraints and increased to 6.91 seconds for 2000 constraints.
Thus, in this setting, it was relatively insensitive to the number of constraints.
By contrast, the optimization time increased substantially as the number of constraints increased.

For BLOBS-3, the Gr\"{o}bner basis computation time increased from 4.10 seconds for 100 constraints to 16.55 seconds for 2000 constraints.
Thus, in settings with more clusters, Gr\"{o}bner basis computation may become a non-negligible bottleneck.
However, the optimization time increased much more sharply, from 77.57 to 7673.87 seconds.
These results further indicate that the optimization algorithm used in the assignment step warrants further investigation.

\begin{table}[t]
\centering
\caption{Runtime breakdown of the proposed algorithm: Gr\"{o}bner basis computation and optimization (in seconds)}
\label{Tab: runtime for proposed}
\small
\begin{tabular}{lrrrr}
\toprule
& \multicolumn{2}{c}{BLOBS-2}
& \multicolumn{2}{c}{BLOBS-3} \\
\cmidrule(lr){2-3}
\cmidrule(lr){4-5}
Number of constraints
& Gr\"{o}bner basis
& Optimization
& Gr\"{o}bner basis
& Optimization \\
\midrule
100  & 4.08 & 7.46  & 4.10  & 77.57   \\
200  & 4.10 & 12.41 & 4.70  & 151.72  \\
500  & 4.16 & 25.78 & 7.44  & 351.86  \\
1000 & 4.27 & 49.05 & 9.19  & 808.76  \\
2000 & 6.91 & 96.25 & 16.55 & 7673.87 \\
\bottomrule
\end{tabular}
\end{table}

\section{Sensitivity to the number of initializations}
\label{Appendix initialization sensitivity}
In the main experiments, each algorithm was run with 100 initializations.
To examine the dependence of the results on this setting, we conducted an additional experiment in which the number of initializations was varied.
Specifically, we used BLOBS-2 and fixed the number of constraints at 100.
The numbers of initializations were set to 100, 200, 500, 1000, and 2000.
The other experimental settings followed those of the main experiments.
In this experiment, runtime was defined as the total CPU time for the specified number of initializations and therefore varied with the number of initializations even for ordinary k-means.
Accordingly, unlike in the main experiments, we report the results for ordinary k-means at each number of initializations.

\begin{figure}[t]
    \centering
    \includegraphics[width=0.95\linewidth]{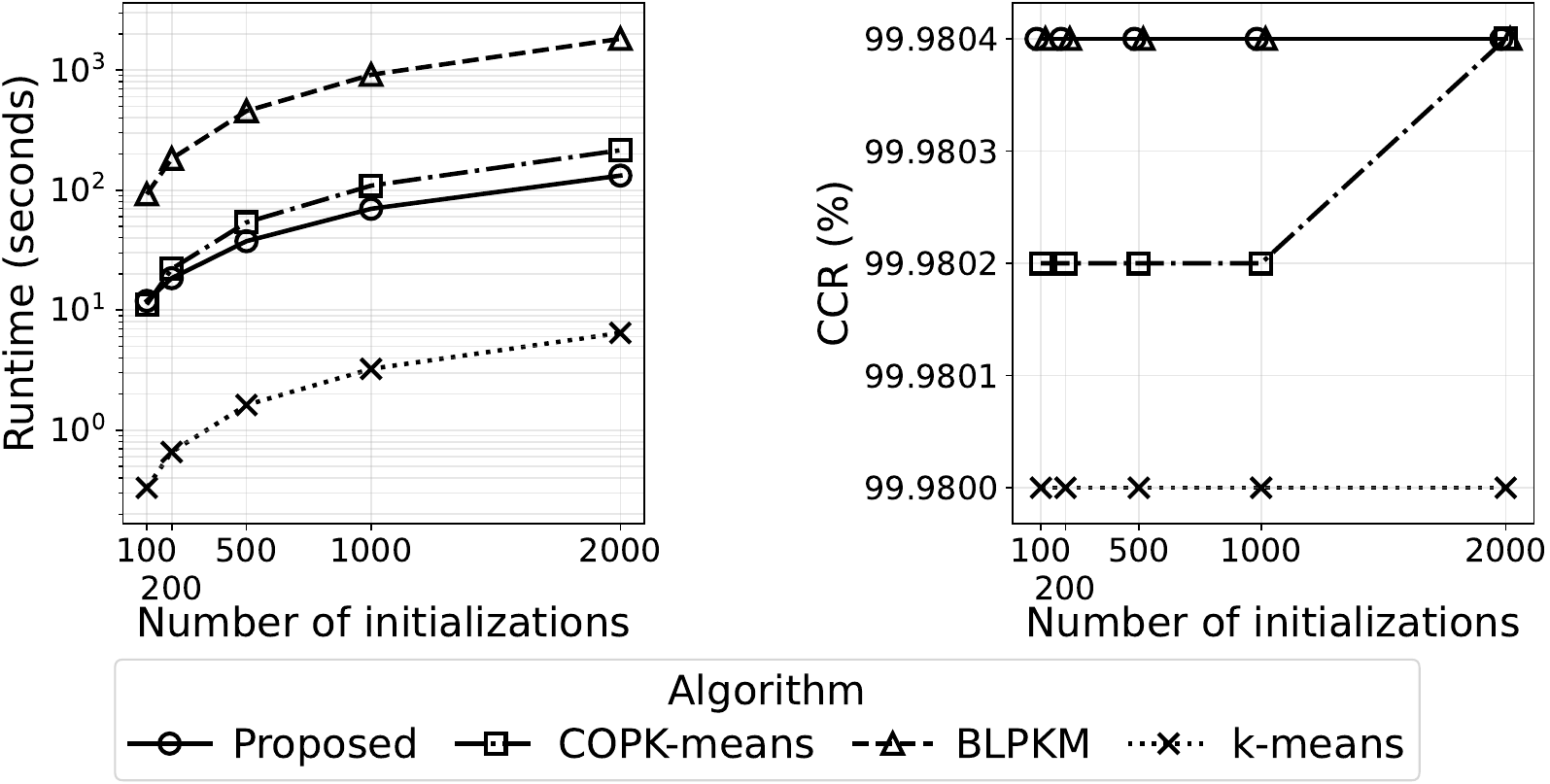}
    \caption{Runtime and Correct Classification Rate (CCR) for different numbers of initializations}
    \label{fig:res:number of initializations}
\end{figure}

The results are shown in Figure~\ref{fig:res:number of initializations}.
Overall, the proposed algorithm was faster than the other constrained algorithms.
This result suggests that the proposed algorithm may be effective when a large number of initializations is required, such as when many local optima are expected or when high cluster estimation quality is required.

By contrast, the CCRs showed little variation as the number of initializations increased.
This result suggests that BLOBS-2 had few local optima and that approximately 100 initializations were sufficient to obtain the best solution in this experiment.

\bibliographystyle{spbasic}
\bibliography{references}

\section*{Declarations}

\subsection*{Funding}
The authors gratefully acknowledge JSPS KAKENHI Grants (JP23K28042, 26K02613).

\subsection*{Competing Interests}
The authors have no relevant financial or non-financial interests to disclose.

\subsection*{Data Availability}
The synthetic datasets used in the experiments can be reproduced using the data-generation code provided by the authors. 
The code is available\\(https://github.com/Hiroki-Oohori-stat/algebraic-constrained-kmeans). 

The real-world dataset used in the experiments, the Mushroom dataset, is publicly available from the UCI Machine Learning Repository \\(https://doi.org/10.24432/C5959T).

\end{document}